\documentclass[
    final            % use final for the camera ready runs
  ]
  {aipproc}

\layoutstyle{8x11single}

\def\ergs{{erg~cm$^{-2}$s$^{-1}$~}}
\def\erg{{erg~s$^{-1}$~}}
\def\fx {f$_{X(0.5-2~\rm{keV})}$}
\def\simlt{\ \raise -2.truept\hbox{\rlap{\hbox{$\sim$}}\raise5.truept   %
\hbox{$<$}\ }}                                                          %
\def\simgt{\ \raise -2.truept\hbox{\rlap{\hbox{$\sim$}}\raise5.truept   %
\hbox{$>$}\ }}
\begin{document}

\title{Blazar surveys with WMAP and Swift}


\author{P. Giommi}{
  address={ASI Science Data Center, ASDC, ESRIN, Frascati, Italy}
,altaddress={On behalf of the Swift blazar team}
}
\author{M. Capalbi}{
  address={ASI Science Data Center, ASDC, ESRIN, Frascati, Italy}
,altaddress={On behalf of the Swift blazar team}
}
\author{E. Cavazzuti}{
  address={ASI Science Data Center, ASDC, ESRIN, Frascati, Italy}
}
\author{S. Colafrancesco}{
  address={ASI Science Data Center, ASDC, ESRIN, Frascati, Italy}
}
\author{S. Cutini}{
  address={ASI Science Data Center, ASDC, ESRIN, Frascati, Italy}
}
\author{\\D. Gasparrini}{
  address={ASI Science Data Center, ASDC, ESRIN, Frascati, Italy}
}
\author{E. Massaro}{
  address={ASI Science Data Center, ASDC, ESRIN, Frascati, Italy}
,altaddress={Dept. of Physics, Sapienza Univ., Rome, Italy}
}
\author{P. Padovani}{
  address={European Southern Observatory, ESO}
}
\author{M. Perri}{
  address={ASI Science Data Center, ASDC, ESRIN, Frascati, Italy}
,altaddress={On behalf of the Swift blazar team}
}
\author{S. Puccetti}{
  address={ASI Science Data Center, ASDC, ESRIN, Frascati, Italy}
}

\begin{abstract}
We present the preliminary results from two new surveys of blazars that have
direct implications on the GLAST detection of extragalactic sources from two
different perspectives: microwave selection and a combined deep X-ray/radio
selection.
The first one is a 41~GHz flux-limited sample extracted from the WMAP 3-year
catalog of microwave point sources.
This is a statistically well defined sample of about 200 blazars and radio
galaxies, most of which are expected to be detected by GLAST.
The second one is a new deep survey of Blazars selected among the radio sources
that are  spatially coincident with serendipitous sources detected in deep X-ray
images (0.3-10 keV) centered on the Gamma Ray Bursts (GRB) discovered by the Swift
satellite.
This sample is particularly interesting from a statistical viewpoint since a)
it is unbiased as GRBs explode at random positions in the sky, b) it
is very deep in the X-ray band (\fx \simgt $10^{-15}$ \ergs) with
a position accuracy of a few arc-seconds, c) it will cover a fairly  large
(20-30 square deg.) area of sky, d) it includes all blazars with radio flux (1.4 GHz)
larger than 10 mJy, making it approximately two orders of magnitude deeper
than the WMAP sample and about one order of magnitude deeper than the deepest existing
complete samples of radio selected blazars, and e) it can be used to estimate the
amount of unresolved GLAST high latitude gamma-ray background and its anisotropy spectrum.
\end{abstract}

\maketitle

\section{Introduction}
It is widely expected that GLAST will detect a large number (probably between 3,000 and
10,000) of extragalactic sources, most of which will be identified as blazars or radio
galaxies. This very rich sample will be used to study the gamma-ray properties of blazars
in great detail. However, in order to secure good samples of candidate gamma-ray sources
and to understand the multi-frequency behavior of blazars from a statistical viewpoint,
it is very useful to create sizable and well defined samples of blazars discovered in
other wave-bands (see, for instance, the $Roma-BZCAT$, Massaro et al., these
proceedings). In this contribution we present the preliminary results from two surveys
that approach this issue from two different perspectives: i) the WMAP 3-year 1~Jy sample
of blazars selected at 41 GHz which defines a microwave selected sample;
ii) the survey of serendipitous blazars detected in Swift X-ray images centered on GRBs, which
defines a combined X-ray/radio selected sample of faint (f$_r >$ 10 mJy at 1.4 GHz,
\fx \simgt $10^{-15}$ \ergs ) blazars.
\section{The WMAP 3-yr 1~Jy  sample of blazars at 41 GHz}
\begin{figure}[!h]
\includegraphics[height=8.5cm,angle=-90]{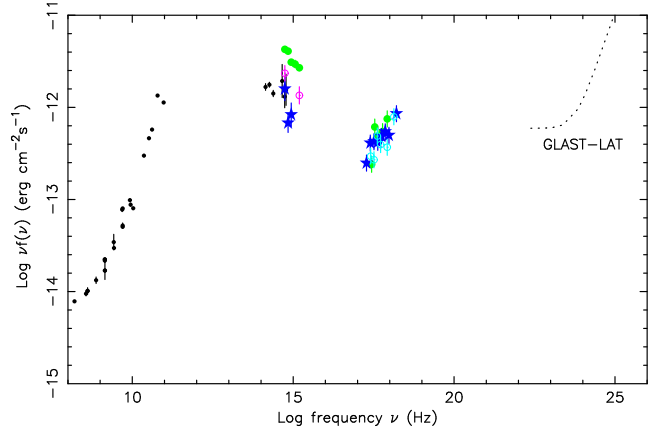}
\caption{The SED of the Blazar WMAP~133=BZQJ0808-0751=PKS~0805-07}
 \label{fig.wmap_bl}
\end{figure}

The catalog of bright point-like sources detected in the WMAP 3-yr data includes 323
objects \cite{Hinshaw06} with microwave fluxes larger than $\approx$ 0.5 Jy. However,
close to the lowest flux level the sample is largely incomplete. We selected a subsample
of 219 sources with 41 GHz flux larger than 1 Jy, that, based on the observed flux
distribution, can be considered complete above this flux level and is therefore best
suited for statistical studies. Since all these sources are also bright at cm wavelength
the large majority of them are well studied radio sources and a good amount of
information about their nature is available in the literature.
To identify all objects in the sample we checked several web-based archives and
literature services and found the following class distribution: FSRQ (156 sources),
BL Lacs (29), SSRQ (9), radio galaxies (17), Staburst galaxies (2), planetary nebulae
(1) and 5 still unidentified objects.
The final sample will be published in the near future and the Spectral Energy
Distribution (SED) of all sources will be made available on the web. As an example Fig.
\ref{fig.wmap_bl} shows the SED of the blazar WMAP~133 built using data from the
literature, WMAP data and optical/UV and X-ray data from Swift satellite (\cite{gio06c}).
The GLAST sensitivity is also plotted to show that blazars so bright at microwave
frequencies can easily detected by GLAST.
\section{The Swift XRT GRB--deep fields}
\begin{figure}[!h]
\includegraphics[height=7.0cm,angle=-90]{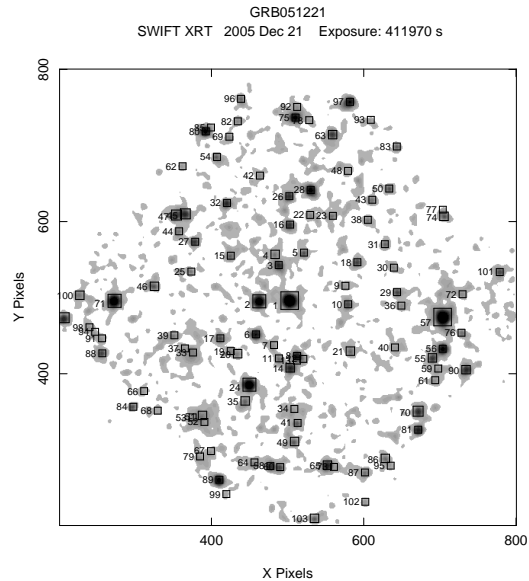}
\caption{The Swift XRT field centered on GRB 051221 with a total exposure of 410,000 seconds.}
 \label{fig.grbdeep}
\end{figure}

\begin{figure}[!h]
\hbox{
 \hspace{-2.0cm}
 \includegraphics[width=4.5cm,angle=-90]{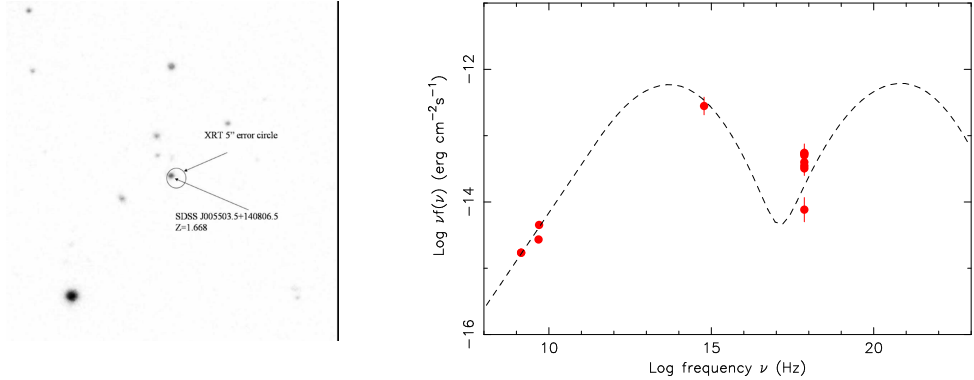}
 \hspace{1.0cm}
 \includegraphics[width=5.0cm,angle=-90]{sed_J0157p0854.ps}

 }
 \caption{Left: the Swift XRT (5 arc-sec) error circle and the optical counterpart of the source SWIFT-XRT~J005503+1408.0.
 Right: the multi-$\nu$ SED of the source SWIFT-XRT~J015700+0854.0.
}
 \label{fig.swift_bl}
\end{figure}

As of today, Swift has discovered over 200 Gamma Ray Bursts (GRBs), a large fraction of
which have been observed with its X-ray (XRT) and Optical-UV (UVOT) telescopes, to monitor the
afterglow for several days or weeks. This resulted in the accumulation of X-ray images that when co-added
reach exposures ranging from $\approx 30,000 $ to over one million seconds, and sensitivity
reaching  $ \approx 10^{-15}$ \ergs in the soft (0.5-2.0 keV) X-ray band (e.g. see Fig. \ref{fig.grbdeep}).
These X-ray images are particularly well suited for statistical purposes since they make an unbiased survey, as GRBs explode at random position
in the sky, are very deep in the X-ray band (\fx \simgt 10$^{-15}$ \ergs).
We have so far analyzed in a standard way the co-added XRT images of 130 GRBs with
exposures ranging from 40,000 to over 600,000 seconds. Within the 90 fields that are at
high Galactic latitude ($|b| > 20$ deg) we have detected over 7,000 point-like
sources, 164 of which coincide (within 10 arc-seconds) with a radio source in the NVSS,
FIRST or SUMSS survey.
From a cross-correlation of this sample 164 sources with catalogs of known objects
we found that none of them was previously known. To start the optical identification
program we used the spectra from the Sloan Digital Sky Survey \cite{sdssdr5}.
As an example Fig. \ref{fig.swift_bl} (left) shows the SDSS image of the field centered on SWIFT-XRT~J005503+1408.0:
only one optical candidate is present within the 5 arc-seconds XRT error radius. Table \ref{tab:a} lists the eight objects
for which we have found a spectrum in the SDSS database. All source are extragalactic with redshift ranging from 0.133 to 1.668.
Figure \ref{fig.swift_bl} (right) shows the SED of another blazar candidate (SWIFT-XRT~J015700+0854.0)
which clearly shows an energy distribution consistent with that of a blazar.
\begin{table}[!h]
\begin{tabular}{lllccc}
\hline
   \tablehead{1}{l}{b}{Source Name \\SWIFT-XRT}
  & \tablehead{1}{c}{b}{SDSS Name \\ }
  & \tablehead{1}{c}{b}{X-ray flux \\0.5-2.0 keV, cgs}
  & \tablehead{1}{c}{b}{X-ray flux \\2.0-10.0 keV, cgs}
  & \tablehead{1}{c}{b}{Radio flux\\ mJy,1.4GHz}
  & \tablehead{1}{c}{b}{Redshift}  \\
\hline
J005503+1408.0 & SDSS J005503.5+140806.5 & $4.6\times 10^{-14}$ & $1.1\times 10^{-13}$&100& 1.668\\
J005514+1407.4 & SDSS J005514.7+140727.0 &$2.4\times 10^{-14}$ & $5.7\times 10^{-14}$&20 & 0.375\\ %\tablenote{Photometric redshift}\\
J082040+3157.1 & SDSS J082040.9+315710.2 & $4.0\times 10^{-15}$ & $9.5\times 10^{-15}$&5 & 0.172\\
J101727+4329.0 & SDSS J101727.5+432904.8 & $5.9\times 10^{-14}$ & $1.3\times 10^{-13}$&197 & 1.173\\
%J114552+5953.3 & $x.x\times 10^{-14}$ & 18&SDSS J114553.0+595320.0 & 0.146\\
J120207+1058.9 & SDSS J120207.9+105851.7 & $1.5\times 10^{-15}$ & $3.0\times 10^{-15}$&21& 0.204\\
J121558+3543.0 & SDSS J121559.3+354302.5 &$5.3\times 10^{-15}$ & $1.2\times 10^{-14}$ &55& 0.133\\
J121559+3521.3 & SDSS J121600.0+352122.0 &$7.9\times 10^{-14}$ & $1.7\times 10^{-13}$&21 & 1.320\\
J151325+3057.9 & SDSS J151326.6+305811.8 &$1.3\times 10^{-15}$ & $3.0\times 10^{-15}$&13 & 0.420\\
\hline
\end{tabular}
\caption{List of X-ray/radio sources with SDSS spectra}
\label{tab:a}
\end{table}
\section{Conclusions and future prospects}
We have presented the preliminary results from two statistically complete and unbiased
surveys of blazars that can be used to:
a)~determine the Gamma-ray properties of well defined samples blazars when the GLAST
survey data will be available, and
b)~determine the statistical properties of blazars over a wide dynamical range, including
a precise estimate of the contribution of blazars to the unresolved gamma-ray background.
We plan to publish the final results on the WMAP 3-year sample before the launch of GLAST and
more complete results on a sample of serendipitous blazars found in about 200 GRB fields by
the end of 2007. Our medium term goal is to use our surveys to obtain $i$)~the radio LogN-LogS of Blazars with fluxes down to 10 mJy,
$ii$)~the radio luminosity function down to $L_{\rm{5GHz}} = 5\times 10^{30}$ \erg Hz$^{-1}$ and its cosmological evolution,
$iii$)~a volume limited X-ray survey out to $z= 0.5$ and $L_X =  5\times 10^{42}$ \erg, and
$iv$)~tight constraints for the contributions of Blazars to the Cosmic Microwave Background and to the
extragalactic gamma-ray background.

%%%%%%%%%%%%%%%%%%%%%%%%%%%%%%%%%%%%%%%%%%%%

\end{document}